# Spectroscopic Signature of Oxidized Oxygen States in Peroxides

*Zengqing Zhuo[1,2,†], Chaitanya Das Pemmaraju[3,†], John Vinson[4], Chunjing Jia[3], Brian Moritz[3], Ilkyu Lee[3], Shawn Sallies[2], Qinghao Li[2], Jinpeng Wu[2], Kehua Dai[2], Yi-de Chuang[2], Zahid Hussain[2], Feng Pan[1]\*, Thomas P. Devereaux[3]\*, and Wanli Yang[2]\**

[1] School of Advanced Materials, Peking University, Shenzhen Graduate School, Shenzhen 518055, People's Republic of China

[2] Advanced Light Source, Lawrence Berkeley National Laboratory, 1 Cyclotron Road, Berkeley CA 94720, United States

[3] Stanford Institute for Materials and Energy Sciences, Stanford University and SLAC National Accelerator Laboratory, Menlo Park, CA 94025, United States

[4] Material Measurement Laboratory, National Institute of Standards and Technology, Gaithersburg, MD 20899, United States

ABSTRACT.

Recent debates on the oxygen redox behaviors in battery electrodes have triggered a pressing demand for the reliable detection and understanding of non-divalent oxygen states beyond conventional absorption spectroscopy. Here, enabled by high-efficiency mapping of resonant inelastic X-ray scattering (mRIXS) coupled with first-principles calculations, we report distinct mRIXS features of the oxygen states in $Li_2O$, $Li_2CO_3$, and especially, $Li_2O_2$, which are successfully reproduced and interpreted theoretically. mRIXS signals are dominated by valence-band decays in $Li_2O$ and $Li_2CO_3$. However, the oxidized oxygen in $Li_2O_2$ leads to partially unoccupied O-$2p$ states that yield a specific intra-band excitonic feature in mRIXS. Such a feature displays a specific emission energy in mRIXS, which disentangles the oxidized oxygen states from the dominating transition-metal/oxygen hybridization features in absorption spectroscopy, thus providing critical hints for both detecting and understanding the oxygen redox reactions in transition-metal oxide based battery materials.

Lithium peroxide, $Li_2O_2$, has been an intriguing system for both structural and chemical properties related to its special oxygen states[1]. Technologically, $Li_2O_2$ is an important air purification agent in spacecraft because it is not hygroscopic as other peroxides and highly reactive with $CO_2$. Recently, the non-divalent oxygen state has attracted increased attention in electrochemical energy storage systems including both alkali-ion batteries and Li-air batteries[2-3]. $Li_2O_2$ is one of the key reaction products in Li-air batteries[2]. $Li_2O_2$, together with $Li_2O$ and $Li_2CO_3$, also dominates the inorganic components of the critical solid-electrolyte-interphase layer formed on negative electrodes of Li-ion batteries[4-6]. More importantly, peroxides may be involved in redox reactions in transition-metal (TM) oxide based electrodes[3,7]. A view which is challenged by other models[8-9].

The redox-active oxygen is a critical concept because conventional batteries rely on only TM redox reactions, and oxygen redox is potentially useful for improving the capacity and energy density of batteries[10-11]. Additionally, oxygen redox could also impact the conceptual developments of catalytic materials[10]. However, although it is widely believed now that oxygen in TM oxides based electrodes could be oxidized to non-divalent states during the electrochemical cycling, as indicated by various core-level X-ray spectroscopy and recent Compton scattering experiments[11-12], the nature of such oxygen redox state has been under active debate[7-9], and the oxidized non-divalent oxygen remains to be understood and reliably characterized[11, 13]. Therefore, a reliable and direct detection of the intrinsic oxygen state, as well as its theoretical understanding, has become crucial for both the fundamental understanding and practical developments of various electrochemical materials.

The challenge of detecting the unconventional oxygen states in TM oxides and the need for better characterizations stem from the fact that conventional O $K$-edge soft X-ray absorption spectroscopy (sXAS) involves entangled contributions through hybridizations between TMs and oxygen. To be more specific, O-$K$ sXAS studies have shown that $Li_2O_2$ displays a characteristic broad feature around 530 eV to 532 eV in sXAS[14]. Unfortunately, this broad feature is located in the same energy range where TM contributes significantly to the O-$K$ sXAS "pre-edge" signals through hybridizations, as identified in the seminal work by de Groot et al. in 1989[15]. Moreover, the overall broadening of the XAS lineshape, due to the presence of a strong core hole created via absorption, often masks low-energy features that are relevant for understanding oxygen redox, complicating a simple interpretation. Indeed, we have recently clarified that most of the claims and conclusions on oxygen redox states based on sXAS experiments merely represent the change of TM states upon electrochemical cycling[13]. An advanced characterization beyond conventional O-$K$ sXAS with better elemental and chemical-bond sensitivity, as well as deeper probe depth, is urgently needed in order to detect and understand the intrinsic nature of oxygen states.

We have recently shown that high-efficiency full energy range mapping of resonant X-ray inelastic scattering (mRIXS) can successfully decipher the entangled O-K signals through the new dimension of information of emission photon energies[13]. By covering the full excitation energy range of O-*K* sXAS, mRIXS detects the energy distribution curves of the fluorescence photons that is only counted as a single number in sXAS, i.e., mRIXS further resolves the emitted photons along the new dimension of emission energy at each absorption energy. Moreover, mRIXS does not suffer from core-hole broadening as in sXAS because the core-hole is filled in the final state of the RIXS process. mRIXS thus becomes a perfect tool-of-choice for reliable and conclusive studies of novel chemical states that cannot be resolved in sXAS[13, 16-18]. Strikingly, O-*K* mRIXS has revealed a sharp feature of the oxygen redox state in battery electrodes that has been buried in conventional sXAS data, with 531 eV excitation energy and 523 eV to 524 eV emission energy, clearly separated from the TM-O hybridization features at 525 eV emission energy[13, 17-18].

However, while mRIXS has been established as a reliable probe of the critical oxygen states involved in the battery electrodes with oxygen redox activities, the interpretation of specific O-*K* mRIXS features has not yet been achieved. The experimental results of mRIXS involve complex processes that are related to electron state configurations, electron correlations, and excitations, which are challenging topics in both fundamental physics and spectroscopic simulations. In general, signals in mRIXS could be categorized into three different types of contributions, the elastic line, non-resonant emission signals from the decays of the occupied valence band electrons to the core holes ("emission lines"), and low-energy excitations[13]. While model compounds may not represent directly the same mechanism as in the complex TM oxide systems, a benchmark study with combined experimental and theoretical results becomes crucial for a general identification of the mRIXS observations, which will shed light on the understanding of the unusual oxygen states involved in the intense debates on TM oxide based energy materials.

In this work, we provide a combined experimental and theoretical mRIXS study of $Li_2O_2$, $Li_2O$, and $Li_2CO_3$. Our central goal is to detect and identify the nature of the aforementioned critical O-$K$ mRIXS feature in Lithium peroxide, thus providing benchmarks and guidelines for understanding the O-$K$ mRIXS findings in energy materials. We note that collecting reliable mRIXS data from $Li_2O_2$ is a nontrivial issue due to the typical low count rates of RIXS experiments and the radiation sensitivity of the material[14, 19]. These technical challenges have now been solved through our recently commissioned RIXS system with ultra-high detection efficiency[20-21], and mRIXS results are successfully collected with controlled sample transfer, cooling, and rastering. Furthermore, advanced simulations with the OCEAN package[22-23] are performed and compared directly with experimental results. We are able to identify the origins of the O-*K* mRIXS experimental features in all the three compounds. We found that mRIXS features of $Li_2O$ and $Li_2CO_3$ are dominated by emission lines from the decays of valence-band electrons. However, a unique excitation feature is defined in $Li_2O_2$ which is a spectroscopic signature of non-divalent oxygen states. Since such oxygen states have partially filled oxygen *2p* bands, we found that the role of Coulomb correlations is critical in adjusting spectral weights. Our combined experimental and theoretical results for $Li_2O_2$ reveal and interpret the critical mRIXS feature at 523.5 eV emission energy across the 529 eV to 532.5 eV excitation energy, which represents a characteristic O-*2p* intra-band excitation in peroxide materials. Strikingly, although much broader along excitation energies, this particular mRIXS feature is close to the observations of the sharp oxygen redox feature in TM oxide based battery materials[13, 17-18], suggesting that the oxygen-redox mRIXS feature found in battery electrodes is intrinsically associated with the partially occupied O-*2p* bands in a highly oxidized TM oxide system.

mRIXS of $Li_2O_2$, $Li_2O$, and $Li_2CO_3$ were collected at the high-efficiency iRIXS endstation of BL8.0.1 of Advanced Light Source[20, 24]. $Li_2O_2$ is unstable under air exposure (forming $Li_2CO_3$), heating (decomposes at 450 °C to $Li_2O$), and X-ray excitations (both $Li_2O_2$ and $Li_2CO_3$ decompose to $Li_2O$)[14, 19]. Therefore, despite the high detection efficiency that allows us to collect a full-range mRIXS map in only about 30 minutes, we employed extensive practices on sample transfer[25-26], liquid $N_2$ cooling, and sample scanning to reduce the radiation effects (see **Supplementary Information**). Still, as elaborated below, some radiation effects remain in our $Li_2O_2$ data. However, comparative studies of all the three materials allow us to distinguish the contributions from material degradation. Additionally, sXAS studies show that $Li_2O_2$ slowly become $Li_2O$ under irradiation[14], the distinct mRIXS features reported here indicate that the signals are intrinsic results of different materials.

First-principles simulations of mRIXS were carried out using the OCEAN package[22-23]. Details of the RIXS implementation within OCEAN have been described previously[27-28] and briefly in **Supplementary Information**. Experimentally determined cubic $Li_2O$[29], hexagonal $Li_2O_2$[30-31] and monoclinic $Li_2CO_3$[32] unit cell structures were used in the simulations. Exchange-correlation effects were treated at the LDA+$U$ level with the Hubbard parameter set to $U$=6 eV on O-$2p$ and C -$2p$ states, similar to previous reports[33-34]. Other details on Brillouin zone, core-hole lifetime, photon polarizations, and adjustments of band gaps are available in **Supplementary Information**.

For the purpose of comparisons, we first present the mRIXS experiments and theoretical interpretation of $Li_2O$ and $Li_2CO_3$. Both have a formal valence of $O^{2-}$ and a nominally fully occupied O-$2p$ shell. We then focus on the specific feature of $Li_2O_2$ arising from its partially filled O-$2p$ band.

The experimental mRIXS of Li$_2$O is shown in **Figure 1(a)**. Other than the elastic line, strong features around 525 eV emission energy (horizontal axis) dominate the whole map, but are separated into two overall packets of intensity along excitation energy regimes, 533 eV to 536 eV and 539 eV to 542 eV (vertical axis). The outgoing photon's emission energy is independent of the incident excitation energy, indicating that these are fluorescence-like emission lines from the decay of electrons from valence bands (VBs) to the core holes[13]. Indeed, the energy of this emission line is consistent with the X-ray emission spectra of Li$_2$O[19]. The origin of the two regimes of excitation energies could be understood by comparing with the O-*K* sXAS spectrum (**Figure S1**). Consistent with the previous report[14], there are two broad absorption features observed for Li$_2$O. The excitation energy ranges of the two sXAS features are the same as those of the two mRIXS intensity packets, which naturally explains the two mRIXS portions from sXAS process, i.e., exciting electrons from O *1s* core level to the unoccupied conduction band states that are further resolved by theoretical calculations.

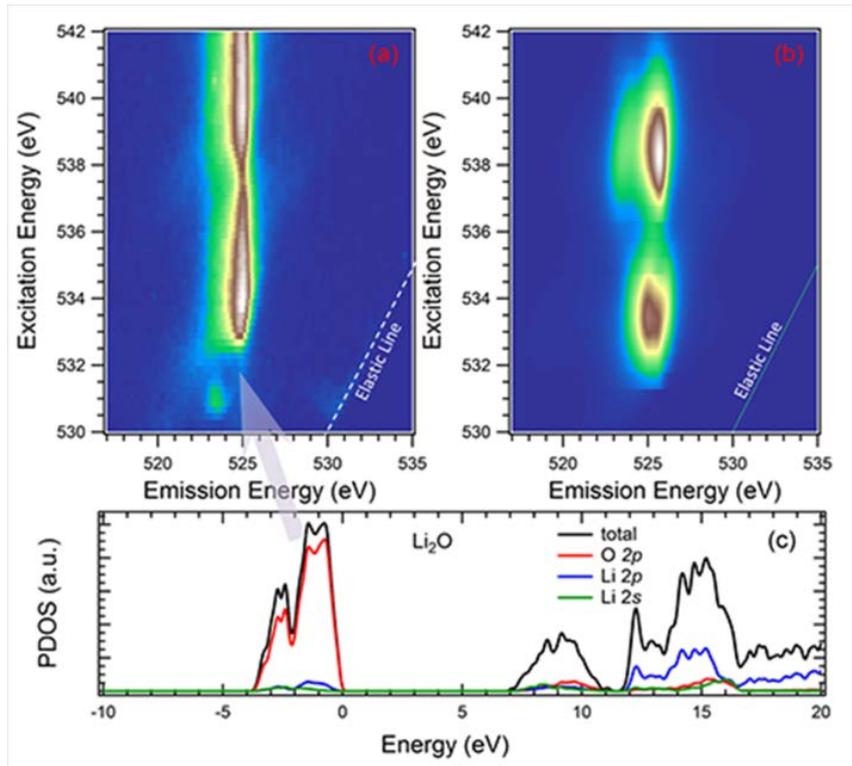

Figure 1. (a) Experimental O *K*-edge mRIXS of $Li_2O$, which is dominated by the emission feature around 525 eV emission energy. Color indicates the intensity distribution of the emitted photons, with blue presenting low intensity and white presenting high intensity. (b) Calculated mRIXS of $Li_2O$, which reproduce the dominate features shown in experimental results. (c) Total and projected density of states of $Li_2O$. The emission line in mRIXS is reproduced in mRIXS calculations by considering the decay of the valence band states, indicated by the arrow.

Simulated mRIXS of $Li_2O$ is displayed in **Figure 1(b)** with calculated total and projected density of states (pDOS) in **Figure 1(c)**. The calculated mRIXS reproduces the dominating features in the experimental results with an emission energy that agrees reasonably well with experiments. The weak feature at the bottom of the mRIXS map (about 530.8 eV excitation energy) is not reproduced in theory but resembles the absorption feature found in peroxides or $O_2$[14, 35], indicating it is likely from impurity. Such a feature in peroxides will be elaborated below. pDOS plots show that O-$2p$ states appears in the conduction band due to the hybridization with Li, sitting around 9 eV and 15 eV above the valence band maximum (VBM). These conduction band states correspond to the two excitation energy ranges around 534 eV and 540 eV in both sXAS (**Figure S1**) and mRIXS (**Figure 1(a)**) experiments. Furthermore, the filled O-$2p$ states in $Li_2O$ form a relatively narrow VB extending over only a 4 eV range below the VBM. Because the O- $p_x$, $p_y$ and $p_z$ orbitals in cubic $Li_2O$ are equivalent, the VB lacks any splitting from anisotropic bonding. This explains the single dominating O-$K$ emission line in mRIXS experiments, which corresponds to the decay from such a narrow VB to the core holes, as also shown in the theoretical mRIXS result in **Figure 1(b)**.

Compared with $Li_2O$, $Li_2CO_3$ is a more complex system and can be considered as a molecular solid with independent carbonate ($CO_3^{2-}$) ions surrounded by $Li^+$ ions. The mRIXS map $Li_2CO_3$ displays two main emission-line (without strong excitation energy dependence) features with several intensity packets in **Figure 2(a)**, centered at 521 eV (low intensity) and 526 eV (high intensity). Again, the main emission lines correspond to decays of VB electrons to the core holes, indicating there are obvious splitting of valence states in $Li_2CO_3$. The well separated islands of mRIXS intensity at 533.7 eV excitation energy are again from sXAS-process, which is directly evidenced by the sXAS peak at the same excitation energy (**Figure S2**) and is known from the C=O bond of carbonates[4-5]. The shift of the weak signals below 532 eV excitation energy in mRIXS is a typical Raman-like shift when excitation energy approaches the absorption edge[36]. The assignments and origins of the observed mRIXS features are further interpreted by mRIXS simulations (**Figure 2(b)**) and the density of states (**Figure 2(c)**). It is clear that C-O hybridization leads to the O-$2p$ pDOS near the bottom of the CB, corresponding to the sXAS feature and mRIXS islands at 533.7 eV excitation energy. Meanwhile, O-$2p$ pDOS near 12 eV result from Li-O hybridization, giving rise to the broad features at higher excitation energies. Compared with $Li_2O$, another major difference of the O $2p$ pDOS of $Li_2CO_3$ is the wide VB distribution over 8 eV range, with many peaks split in two groups separated by a gap of ~1.5 eV. The upper band is almost entirely composed of O-$2p$ states while the lower band is a mix of O-$2p$ and C-$2p$ states (**Figure 2(c)**). The upper and lower VBs lead to two separated emission lines in the calculated mRIXS, centered respectively around 526 eV and 521 eV with the former being more intense than the latter. The split features at different emission energies are also reproduced (**Figure 2(b)**), consistent with experimental results. However, there is quantitative discrepancy between the experimental results and theoretical calculations on the energy values of the emission lines, especially in the lower excitation energy range. This is traced to the choice of the U value in our DFT+U calculations. As shown in **supplementary information (Figure S4)**, although U = 6 eV is the optimal value for describing hole-polaron behavior in previous publications[33-34], decreasing the U value from 6 eV to 3 eV leads to a better quantitative agreement with experiments. In this context, GW[37] quasiparticle corrections to DFT or DFT+U single-particle energies could be relevant to improving the predictive accuracy of first-principles BSE RIXS[28] approach. We also note that other theoretical method should be further explored for RIXS calculations, for example, a very recent theoretical work based on Wannier orbital method could reproduce the experimental results of $Li_2CO_3$.[38]

Therefore, all the mRIXS observations in $Li_2O$ and $Li_2CO_3$ could be simulated and assigned to emission lines corresponding to the decay of VB electrons to the core holes generated during the sXAS process. The mRIXS contrast between the two systems is mainly due to the different VB configurations.

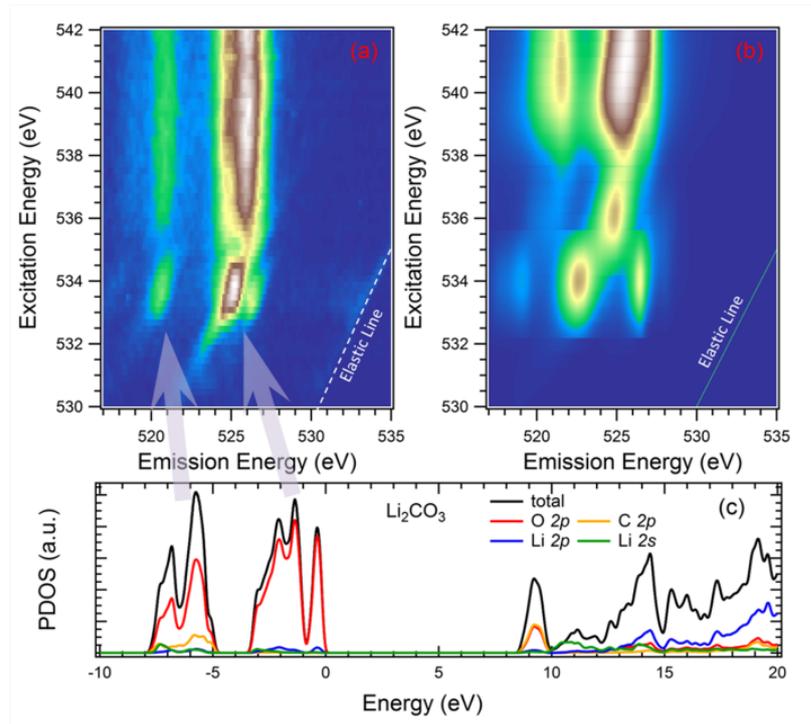

Figure 2. (a) Experimental O *K*-edge mRIXS of $Li_2CO_3$. Two emission features centered at 521 eV and 526 eV emission energy are observed. (b) Calculated mRIXS of $Li_2CO_3$ successfully reproduces the experimental features by considering the emissions from the decay of the split valence band states, which are shown in (c).

**Figure 3** displays the experimental and theoretical mRIXS results of $Li_2O_2$ with a significantly changed DOS configuration due to the partially occupied O *2p* bands. $Li_2O_2$ can be considered as a molecular solid comprised of independent $O_2^{2-}$ peroxide ions surrounded by $Li^+$ ions. With the peroxo bond axis oriented along the c-axis of the $Li_2O_2$ crystal, bonding is highly anisotropic with O *2p* states bifurcating into distinct π/π* ($p_x$, $p_y$) and σ/σ* ($p_z$) bonding/anti-bonding groups that are well separated in energy as shown in **Figure 3(c)**. The conduction band minimum in $Li_2O_2$ is essentially made up of unoccupied $p_z$ orbitals oriented along the peroxo bond in $\sigma_u^*$ symmetry.

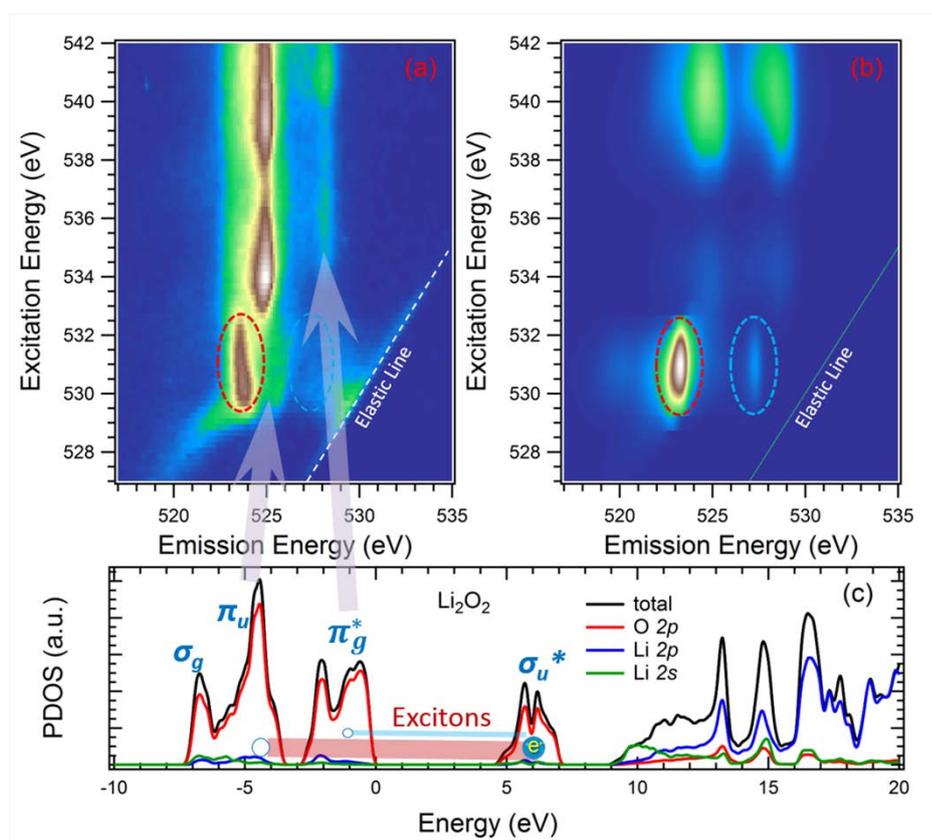

Figure 3. (a) Experimental O $K$-edge mRIXS of $Li_2O_2$. A specific feature centered at 523.7ev and two emission features at 525 eV and 528 eV emission energy are observed. (b) Calculated mRIXS of $Li_2O_2$ reproduces the experimental results with the striking feature at 523.7 eV emission energy (red circle). (c) Total and projected density of states of $Li_2O_2$. Decay of the split valence band states lead to the two main

emission features centered at 525 eV and 528 eV. However, the specific feature centered at 523.7 eV emission energy is from the excitations between the occupied and unoccupied O-$2p$ states in the vicinity of the Fermi Level, due to the partially occupied O-$2p$ states in peroxides.

Like $Li_2CO_3$, the O $2p$ pDOS in the VBs of $Li_2O_2$ is distributed over a wide energy range and splits into regions with $\sigma_g$, $\pi_u$ and $\pi_g$* character with a gap of ~1eV between the $\pi_u$ and $\pi_g$* states (**Figure 3(c)**). Therefore, two mRIXS emission-line features arise from the decay of electrons in the ($\sigma_g$, $\pi_u$) and $\pi_g$* states to the core holes, leading to the two vertical mRIXS features at 525 and 528 eV emission energies (**Figure 3(a)**). Calculations based on VB decay again reproduce these emission-line features (**Figure 3(b)**). The feature at 525 eV emission energy is attributed to decays from the lower energy $\pi_u$ states, while the 528 eV feature is attributed to decays from the $\pi_g$* states. The excitation energy dependence of the mRIXS features are again consistent with the sXAS results (**Figure S3**), where broad features at 529 eV to 533.5 eV, 533 eV to 536eV, and above 538 eV are observed. The low excitation energy 529 eV to 533.5 eV sXAS feature corresponds to the special O-O bonding in $Li_2O_2$[13], i.e., the $\sigma_u$* states from unoccupied $p_z$ orbitals as explained above. Features above 538 eV excitation energy are from sXAS process to the high-energy Li-O hybridization states (**Figure 3(c)**). However, the broad feature in the intermediate excitation energies, 533 eV to 536 eV, has no corresponding pDOS, thus cannot be reproduced from theoretical calculations, but it matches almost exactly the strong feature of $Li_2O$ (**Figure 1**). Our previous study has shown that $Li_2O_2$ could be degraded into $Li_2O$ under soft X-ray exposure[14]. We therefore assign the signals at 533 eV to 536 eV excitation energies to $Li_2O$ from irradiation effect and/or impurity, even with our controlled and fast experimental scans.

The most striking finding of $Li_2O_2$ mRIXS is the intense feature near the 529 eV to 533.5 eV excitation and 523.7 eV emission energies, which appears as the strongest mRIXS feature in theoretical calculations (**Figure 3(b)**). As mentioned above, the sXAS signals at this energy range corresponds to the unoccupied $\sigma_u^*$ states from the O-O bonding in peroxides[14]. However, the emission energy of this specific feature, 523.7 eV, is obviously different from the emission-line features (525 and 528 eV for $Li_2O_2$), indicating a different spectroscopic origin. More importantly, although with different broadening levels in excitation energy, the emission energy of this feature covers the oxygen redox mRIXS feature found in the battery electrodes with oxidized oxygen[13, 17-18]. It is therefore instructive and critical to analyze the character of this striking feature.

**Figure 4(a) and (b)** show the density isosurface plots of the electron and hole contributions in our calculations that reproduces this 523.7 eV emission feature in theory. We choose the incoming photon polarization along the peroxo bond ($p_z$) direction and the outgoing polarization to be perpendicular to it, along $p_x$. Based on the incoming and outgoing photon energies, this mRIXS feature is reproduced successfully through a specific excitonic state, where the electron and hole has $\sigma_u^*$ and $\pi_u$ characters, respectively. As directly shown in **Figure 4**, the electron part of the excitonic wavefunction is composed of orbitals oriented along the peroxo bond axis, reflecting its $p_z$ derived antibonding σ* character. The hole density is formed predominantly from orbitals oriented along $p_x$ orbitals with π bonding character. Therefore, the critical mRIXS feature at 523.7 eV emission and 529 eV to 533.5 eV excitation energy is an O-*2p* intra-band excitation between the occupied π bonding states and the unoccupied σ* antibonding states (**Figure 3c**). The mismatch between the energy difference of the excitonic states (**Figure 3c**) and experimental energy loss is mostly due to the core hole effect, which is accounted for in theoretical calculations that show consistent results to experiments (**Figure 3b**). We note that the mRIXS feature of $Li_2O_2$ is broader than the calculation results, and the O-*K* sXAS also shows a relatively broader peak compared with hard X-ray results (**Figure S3**)[30], therefore surface degradation/contamination of the $Li_2O_2$ at least partially contribute to the broadening of this particular mRIXS feature. Dynamic disorders due to finite temperature may also contribute to the experimental broadening. It is important to note that, compared with the mRIXS feature of $Li_2O_2$, the much sharper feature around the same emission energy in TM oxide based battery electrodes does imply differences in excitations and/or associated electron states, a topic that deserves further studies to clarify.

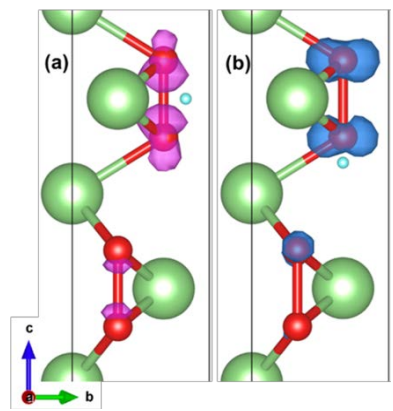

Figure 4. Density iso-surface plots of the electron (a) and hole (b) contributions within the excitonic wavefunction of the RIXS final state corresponding to the 523.7 eV emission mRIXS feature of $Li_2O_2$. The blue (purple) shapes indicate the distributions of the holes (electrons) for plotting the electronic (hole) part of the two particle wavefunction. Green and red balls represent the lattice Li and O, respectively.

In summary, the mRIXS comparisons between the $Li_2O_2$ and $Li_2O/Li_2CO_3$ systems show clear spectroscopic differences between the response of (oxidized) O-*2p* in peroxides and the fully occupied *2p* orbitals of the $O^{2-}$ states. Theoretical calculations not only reproduce the mRIXS features, they also clarify that the specific $Li_2O_2$ mRIXS signals at 523.7 eV emission energy originates from O-*2p* intra-band excitations, providing a spectroscopic signature for studying non-divalent oxygen state. The results suggest that the mRIXS feature found in battery electrodes with similar emission energy, but sharper excitation distribution[13, 17-18], is not a feature from the decay of the occupied valence band electrons as claimed in previous publications. Instead, it indicates a specific excitation in highly oxidized TM oxide systems. However, it is important to note that one should not simply take this work as evidences of peroxides in battery electrodes. The excitations revealed here are inherent to highly oxidized systems, e.g., peroxides, superoxides, and even $O_2$ gas should all display excitonic features alike. Although no full-energy-range mRIXS results have been reported for other non-divalent oxygen compounds, a RIXS single spectrum of $O_2$ gas collected with 530.8 eV excitation energy did indicate a feature at 523.7 eV emission energy[39]. As discussed above, further works are still necessary to clarify the exact excitations responsible for the sharp mRIXS feature in electrodes based on TM oxides[13, 17-18]. Nonetheless, the results and analysis here conclude that the mRIXS feature at 523.7 eV emission energy emerges from specific excitations in highly oxidized systems, not from the decay of valence band electrons as indicated in previous works on battery electrodes. This clarification provides a critical foundation for further studies of the oxidized oxygen states in the more complex TM oxide systems, especially the electrochemical materials with oxygen redox activities.

**Supporting Information**. Detailed descriptions of XAS and mRIXS experimental process. Detailed descriptions of theoretical calculation methods.

**AUTHOR INFORMATION**

**Corresponding Author**

* FP (panfeng@pkusz.edu.cn); TPD (tpd@stanford.edu); WY (wlyang@lbl.gov)

**Notes**

† Zengqing Zhuo and Chaitanya Das Pemmaraju contributed equally to this work.

The authors declare no competing financial interests.

**ACKNOWLEDGMENTS**

Advanced Light Source is supported by the Director, Office of Science, Office of Basic Energy Sciences, of the U.S. Department of Energy under Contract No. DE-AC02-05CH11231. This work is also supported by Guangdong Innovation Team Project (No. 2013N080) and Shenzhen Science and Technology Research Grant (peacock plan KYPT20141016105435850). Theoretical works are supported by the US Department of Energy, Office of Science, Office of Basic Energy Sciences, Division of Materials Sciences and Engineering, under Contract No. DE-AC02-76SF00515. This research used resources of the National Energy Research Scientific Computing Center (NERSC), a U.S. Department of Energy Office of Science User Facility operated under Contract No. DE-AC02-05CH11231. Additionally, some of the computing for this project was performed on the Sherlock cluster and support from Stanford University and the Stanford Research Computing

Center is acknowledged. WY would like to thank L. Andrew Wray (NYU) for thoughtful discussions.

013110.

Supplementary Information

# Spectroscopic Signature of Oxidized Oxygen States in Peroxides

*Zengqing Zhuo[1,2†], Chaitanya Das Pemmaraju[3†], John Vinson[4], Chunjing Jia[3], Brian Moritz[3], Ilkyu Lee[3], Shawn Sallies[2], Qinghao Li[2], Jinpeng Wu[2], Kehua Dai[2], Yi-de Chuang[2], Zahid Hussain[2], Feng Pan[1]\*, Thomas P. Devereaux[3]\*, and Wanli Yang[2]\**

[1] School of Advanced Materials, Peking University, Shenzhen Graduate School, Shenzhen 518055, People's Republic of China

[2] Advanced Light Source, Lawrence Berkeley National Laboratory, 1 Cyclotron Road, Berkeley CA 94720, United States

[3] Stanford Institute for Materials and Energy Sciences, Stanford University and SLAC National Accelerator Laboratory, Menlo Park, CA 94025, United States

[4] Material Measurement Laboratory, National Institute of Standards and Technology, Gaithersburg, MD 20899, United States

\* FP (panfeng@pkusz.edu.cn); TPD (tpd@stanford.edu); WY (wlyang@lbl.gov)

**Supplementary Methods**

Experimental

Li$_2$O, Li$_2$O$_2$, and Li$_2$CO$_3$ materials were used as received from Sigma-Aldrich\*,

---

\*Certain commercial materials and software are identified in this paper to foster understanding. Such identification does not imply recommendation or endorsement by the National Institute of Standards and Technology, nor does it imply that the materials identified are necessarily the best available for the purpose.

with purity of 97%, 90%, and 99.97% by mass, respectively. The samples were prepared and loaded in a high-purity (99.999% by volume) Ar glove box with water and oxygen concentrations below 1 mg/L (1 ppm). Materials are pressed onto indium foil mounted on sample holders. The samples were transferred into measurement chamber by using a specially designed sample transfer kit to avoid any air exposure[1-2].

sXAS and RIXS measurements were performed in the high-efficiency iRIXS endstation at Beamline 8.0.1 of the Advanced Light Source at Lawrence Berkeley National Laboratory[3]. The beamline undulator and spherical grating monochromator supply a linearly polarized photon beam. The linear polarization of the incident beam is parallel to the scattering plane. The energy resolution of the excitation X-ray beam is about 0.3 eV. The sXAS data shown in this work are total and partial fluorescence yield (TFY and PFY) extracted directly from our mRIXS results[4]. mRIXS data were collected through a high-transmission soft X-ray spectrometer with energy resolution of about 0.3-0.35 eV on emission energy[5]. A series of RIXS spectra have been recorded across the O $K$-edge with steps of 0.2 eV in excited energies. The recorded spectra were then plotted in color scale and combined into a mRIXS map[6]. All the data have been normalized to the beam flux measured by a clean gold mesh upstream of the endstation.

The mRIXS maps shown in this work are accomplished within half an hour. Samples are cooled by Liquid $N_2$ and remain itinerant during the experiments to reduce radiation damage effects. Radiation damage has been carefully monitored before and after RIXS experiments. We note that radiation damage effect cannot be completely ruled out, however, as elaborated in the manuscript, based on our previous studies of irradiation effects of the same materials, which shows the same final formation of $Li_2O$ after extended soft X-ray exposure on one sample spot, the distinct RIXS maps that we focus on in this work directly show they are intrinsic material features.

**Theoretical Calculations**

First-principles simulations of the RIXS spectra of $Li_2O$, $Li_2O_2$ and $Li_2CO_3$ were carried out using the OCEAN package[7-8], which implements the Bethe Salpeter Equation (BSE) formalism for core and valence excitations. Details of the RIXS implementation within OCEAN have been described previously[9-10]. Briefly, OCEAN employs the Kramers-Heisenberg formula for RIXS while treating electron-hole interaction effects in both the core-excited intermediate state and valence-excited final state at the BSE level. The BSE is solved in a two particle electron-hole basis built from occupied and virtual Kohn-Sham (KS) orbitals obtained from a Density Functional Theory (DFT)[11-12] simulation of the ground state. Core orbitals relevant to the RIXS measurement of interest are included in the description via the projector augmented wave technique. Ground state simulations of $Li_2O$, $Li_2O_2$ and $Li_2CO_3$ were carried out using the QUANTUM ESPRESSO[13] code. Norm-conserving pseudopotentials treating the Li:2$s$, O:2$s$,2$p$ and C:2$s$,2$p$ states in the valence were employed in conjunction with a planewave cutoff of 100 Ry. Experimentally determined cubic $Li_2O$[14], hexagonal $Li_2O_2$[15] and monoclinic $Li_2CO_3$[16] unit cell structures were used in the simulations. Brillouin zone integration was carried out employing Γ-centered 8x8x8, 8x8x4 and 6x6x6 k-point grids for $Li_2O$, $Li_2O_2$ and $Li_2CO_3$ respectively.

In order to partially mitigate band gap underestimation and excessive delocalization artifacts[17] inherent to semi-local DFT, exchange-correlation effects were treated at the LDA+$U$[18] level with the Hubbard parameter set to $U$=6 eV on O-2$p$ and C-2$p$ states. Similar parameters were previously employed to model polaronic behavior in $Li_2O_2$ and $Li_2CO_3$[19] as well as other transition metal oxides[20]. Residual bad gap underestimation was corrected for through a simple rigid shift applied to the conduction band states in order to approximate quasiparticle band gaps of 7.4 eV ($Li_2O$)[21], 4.9 eV ($Li_2O_2$) and 8.8 eV ($Li_2CO_3$)[19]. We note in this context that the RIXS simulations are sensitive to the choice of the $U$ parameter employed. Particularly in $Li_2CO_3$, the choice of $U$=6 eV which is in the optimal range for describing hole-polaron behavior in oxides produces some additional shifts in the RIXS emission features relative to experiment (see Figure 2 of main text). A smaller value of $U$=3 eV for the O-2$p$ and C-2$p$ states in $Li_2CO_3$ yields RIXS maps in better agreement with experiment (see figure S4). Kohn-Sham wavefunctions and eigenvalues obtained through this procedure were used in the subsequent BSE calculations. An O K-edge core-hole lifetime parameter of 0.14 eV was used in the simulations. In simulating RIXS spectra incoming and outgoing polarizations were considered to be orthogonal to confirm with the experimental geometry and the spectra were averaged over polarizations along the three Cartesian axes.

In order to facilitate comparison with experiment, the incoming photon energy axis was calibrated by matching the calculated X-ray absorption spectrum onset in each case to experiment. Furthermore, an additional Gaussian broadening of 0.8 eV was applied to the simulated RIXS and XAS along the incoming photon energy axis.

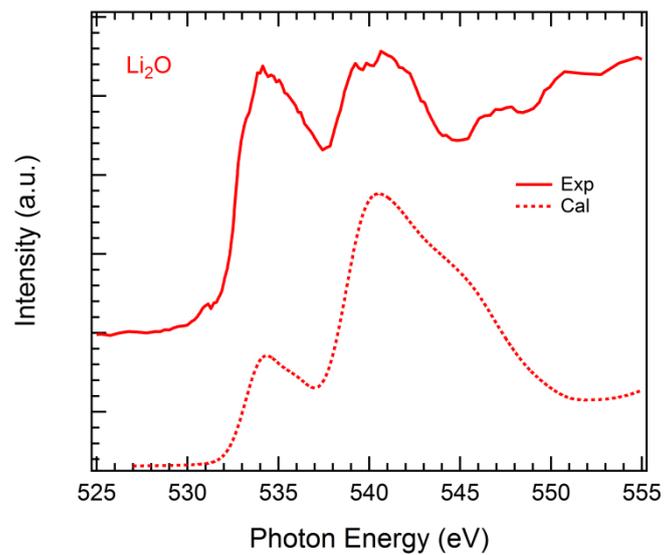

Figure S1 O *K*-edge XAS experimental (solid line) and calculated (dotted line) of Li$_2$O, both of which show two broad feature around 534.4 eV and 541 eV.

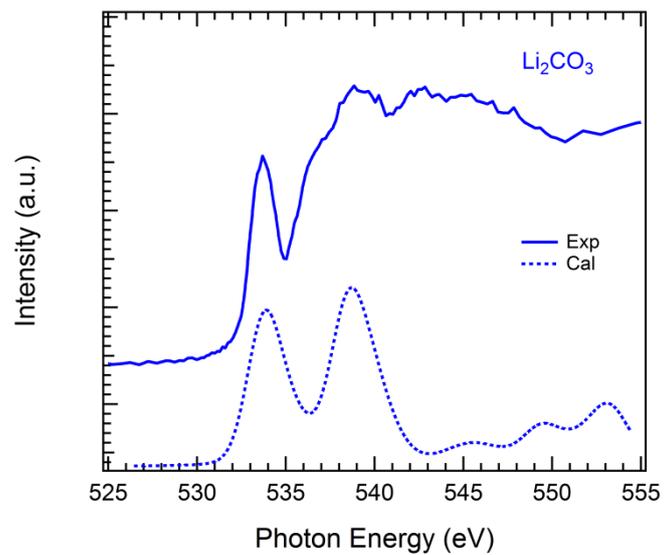

Figure S2 O *K*-edge XAS experimental (solid line) and calculated (dotted line) of Li$_2$CO$_3$. The 533.7 eV feature is related C=O bond of carbonates.

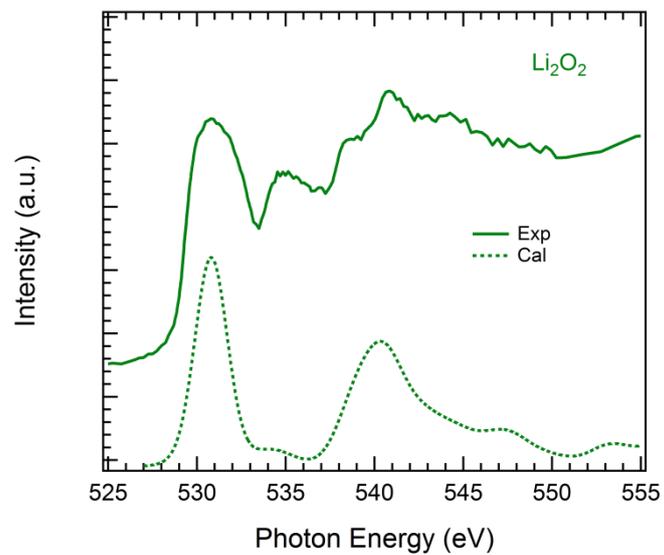

Figure S3 O *K*-edge XAS experimental (solid line) and calculated (dotted line) of Li$_2$CO$_3$. The 529-533.5 eV broad feature corresponds to the special O-O bonding in Li$_2$O$_2$.

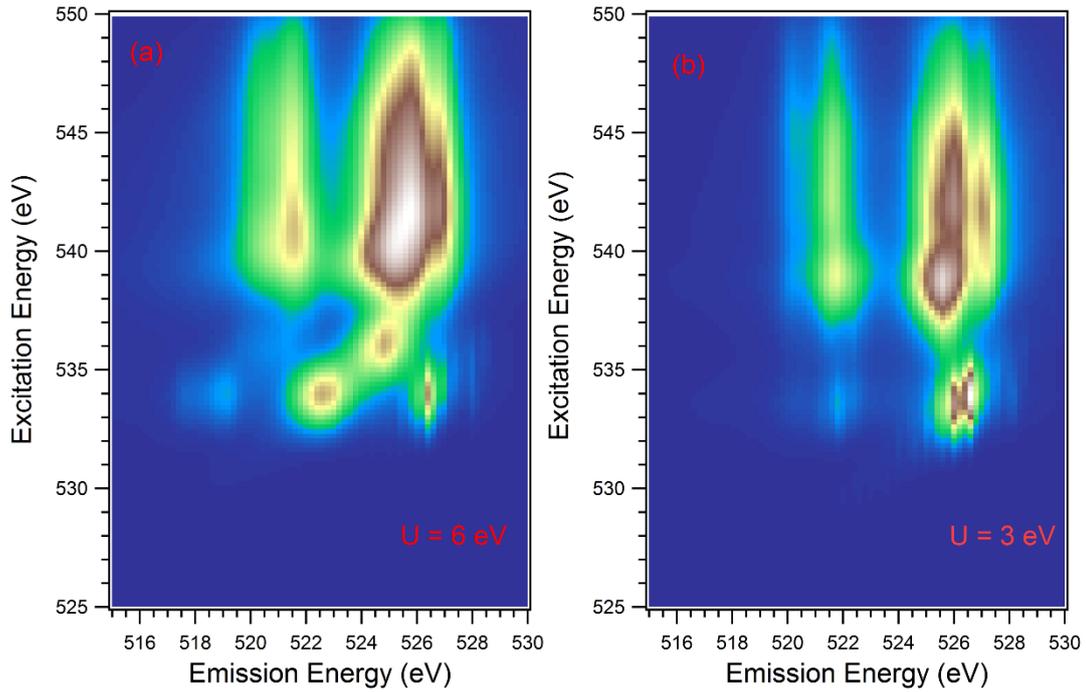

Figure S4 Simulated O *K*-edge mRIXS of $Li_2CO_3$ with two different Hubbard "U" parameter. For U=6 eV, shifts in the emission features corresponding to excitation near 534 eV are observed, whereas mRIXS obtained at U=3 eV displays a different intensity map and is in better agreement with experiments (see Figure 2 of main text). Such a contrast indicates the importance of the U parameter in the quantitative determination of the mRIXS results.